\documentclass[conference]{IEEEtran}
\usepackage{float}
\usepackage{subfig}
\usepackage{amsmath,epsfig,psfrag} 
\usepackage{amsfonts,color,graphics,graphicx}
\usepackage{algorithm} 
\usepackage{algpseudocode}
\usepackage{multirow}
\usepackage{rotating}
\usepackage[numbers]{natbib}
\usepackage[numbers]{natbib}
\usepackage[normalem]{ulem}
\renewcommand{\baselinestretch}{1}

\usepackage{verbatim}

\newcommand\blfootnote[1]{%
  \begingroup
  \renewcommand\thefootnote{}\footnote{#1}%
  \addtocounter{footnote}{-1}%
  \endgroup
}

\begin{document}

\title{Models for Predicting Community-Specific Interest in News Articles}

\author{\IEEEauthorblockN{Benjamin D. Horne}
\IEEEauthorblockA{\textit{Rensselaer Polytechnic Institute}\\
Troy, New York 12180 \\
horneb@rpi.edu}
\and
\IEEEauthorblockN{William Dron}
\IEEEauthorblockA{\textit{Raytheon BBN Technologies}\\
Cambridge, Massachusetts 02138 \\
will.dron@raytheon.com}
\and
\IEEEauthorblockN{Sibel Adal{\i}}
\IEEEauthorblockA{\textit{Rensselaer Polytechnic Institute}\\
Troy, New York 12180\\
adalis@rpi.edu}
}

\maketitle

\begin{abstract}
In this work, we ask two questions: 1. Can we predict the type of community interested in a news article using only features from the article content? and 2. How well do these models generalize over time? To answer these questions, we compute well-studied content-based features on over 60K news articles from 4 communities on reddit.com. We train and test models over three different time periods between 2015 and 2017 to demonstrate which features degrade in performance the most due to  concept drift. Our models can classify news articles into communities with high accuracy, ranging from 0.81 ROC AUC to 1.0 ROC AUC. However, while we can predict the community-specific popularity of news articles with high accuracy, practitioners should approach these models carefully. Predictions are both community-pair dependent and feature group dependent. Moreover, these feature groups generalize over time differently, with some only degrading slightly over time, but others degrading greatly. Therefore, we recommend that community-interest predictions are done in a hierarchical structure, where multiple binary classifiers can be used to separate community pairs, rather than a traditional multi-class model. Second, these models should be retrained over time based on accuracy goals and the availability of training data. \blfootnote{Published at IEEE MILCOM 2018 in Los Angeles, CA, USA}
\end{abstract}


\section{Introduction}
Understanding community engagement ranges from important to crucial in a wide range of military missions. However, doing so in a meaningful and automated way has and continues to be an extremely challenging problem. The military must be aware of micro cultural issues and how populations will react to current events. Provided with this understanding, the military can adapt and spend resources on areas most crucial to tipping population sentiment in our favor. We look to augment the work of anthropologists and sociologists by automating the understanding of distinct communities. To this end, we use Reddit, as an initial source of micro-communities, to detect how communities will react to events as reported by national and local media. Specifically, we ask the question: \textbf{Q1}: Can we predict the type of community interested in a news article using only features from the article content? Prior work has built models to predict general popularity, both with cold-starts (i.e. based only on message content, before the spread of information)~\cite{arapakis2014feasibility, bandari2012pulse} and warm-starts (i.e. based on early popularity)~\cite{lee2010approach, lerman2010using, szabo2010predicting}, however, there has been little work in predicting community-specific popularity of a news article. 

In this work, we approach the problem by building and validating cold-start machine learning models to classify the community of interest among 4 distinct news communities on Reddit. We use well-studied content-based features, used in previous studies for news credibility~\cite{horne2018accessing} and news popularity~\cite{bandari2012pulse}. Further, we examine models using intuitive feature groups to better understand what signals predict best and how they differ between these communities.

One of the many challenges with predicting human activities is the constant evolution of behavior. This often creates a moving target for machine learning models, which can be complex and hard to control for. The problem we address in this paper has many naturally evolving parts: the news cycle, the political climate of a region, and the community preferences. Thus, to explore concept drift in community-specific popularity prediction, we ask a secondary question: \textbf{Q2}: How well do these models generalize over time? To answer this, we gather Reddit news community data from a 3 year time period and emulate a machine learning model that has not been retrained for 2 years. This analysis is done for each feature group to show which features generalize better over time. 

Lastly, this work builds a foundation for several important future works, particularly in a military setting. Since this work focuses on community-specific popular, rather than general popularity, it relates to better understanding targeted or malicious media coverage, where news articles or blogs are written with the intention of provoking certain communities. One of the objectives of our research is to eventually be able to detect such types of non-traditional attacks. Furthermore, the techniques explored in this work can be extended to many more types of communities, whether those are on different media platforms or centered in different countries.

Our results show that we can predict the community-specific popularity of news articles with high accuracy, but practitioners should approach these models carefully. While our models can predict community interest with close to 100\% accuracy, these predictions are both community-pair dependent and feature dependent. In addition, these feature groups generalize over time differently, with some only degrading slightly over 2 years, but others becoming useless. Hence, we make two recommendations: 1. Community-interest predictions should be done with a hierarchical model, where multiple feature-filtered binary classifiers can be used to separate community-interest pairs, rather than a traditional multi-class model, 2. These models should be retrained over time based on the application's accuracy goals and the availability of data. 

\section{Related Work}
There are many prior works on general news popularity. The majority of these works use and develop content-based features to capture popularity, such as headline features~\cite{piotrkowicz2017headlines, reis2015breaking} or body content~\cite{horne2017impact}. However, others have based predictions on other signals such as users comments~\cite{tatar2011predicting} and early popularity~\cite{lee2010approach, lerman2010using, szabo2010predicting}. Even more general, there is a large body of work on content popularity prediction, not focused on news articles. These works include predicting the popularity of comments on Reddit~\cite{horne2017identifying, jaech2015talking}, predicting the popularity of tweets or hashtags on Twitter~\cite{ma2013predicting, zaman2014bayesian}, and predicting the popularity of videos~\cite{li2013popularity, szabo2010predicting}.

While news popularity has been studied extensively, \emph{community-specific} popularity has not. The only prior work attempting to predict community-specific interest is in~\cite{horne2018accessing}, in which a simple content based model is deployed on a 3 month Reddit data set, achieving 77\% accuracy. We hope to both improve upon this model and gain a better understanding of what signals generalize well over time. 

\begin{table*}
\begin{center}
\fontsize{7.95}{8}\selectfont
\hspace*{-0.0in}\begin{tabular}{|c||p{6in}|}
\hline
\textbf{Community} & \textbf{Description} \\
mainstream &  News community that does not allow opinion-based articles or articles that do not properly report a story. (Generally reliable news stories)\\
conspiracy & Conspiracy theory community that does not censor posts and encourages news about unconfirmed hypotheses. (Questionable news stories) \\
bias1 &  News community that is focused on stories from one political viewpoint, that may or may not be misleading (Hyper-partisan news stories)\\
bias2 & News community that is focused on stories from one political viewpoint, opposite of bias1 (Hyper-partisan news stories) \\

\hline
\end{tabular}
\caption{Description of communities used in study.}\label{reddits}
\end{center}
\end{table*}

\section{Data}
In order to address the problem of community-specific popularity, we will use data from a set of news communities on reddit.com. Reddit is a social news-aggregation platform made up of interest-based communities called \textit{subreddits}. Each subreddit has subscribers who can post urls to news articles, comment on posts, and vote for a post, which roughly determines its placement on the page. Further, each subreddit has a moderation team that ensures content meets the community's standards. These standards can vary widely, from requiring informative news to news of a specific view point. This clear structure and diversity of news-based communities provides an ideal setting to explore community-specific interest in news. While this problem can be thought of more generally (not only Reddit communities) and need not be focused on country specific  news communities, we use this data set to provide a clear testing bed for our methods. Further, Reddit is a widely used platform, ranking 6th in global popularity according to alexa.com in 2018. Reddit has also been shown useful in other prediction tasks, such as ranking popular comments in discussions~\cite{horne2017identifying} and general popularity studies~\cite{hessel2017cats, lakkaraju2013s, tran2016characterizing}.

In this study, we use 4 distinct subreddits: a general news community, a conspiracy news community, and two hyper-partisan news communities. We will call these communities \textit{mainstream}, \textit{conspiracy}, \textit{bias1}, and \textit{bias2}, respectively. A description of each community can be found in Table~\ref{reddits}. It is clear that these communities capture diverse types of news in terms of both view point and reliability. Specifically, we expect conspiracy to contain questionable/unconfirmed (conspiracy-theory) news, while both bias1 and bias2 contain news from two extreme view points (opposite viewpoints). These 3 communities contrast with mainstream, which seeks to curate factual/non-opinion based news.

To construct this data set, we first collect posts from each community on Reddit for 3 months in 2017 using the Reddit API\footnote{https://goo.gl/qe2xpJ}. This collection includes the news article url, the post time, the post score (based on community members votes), and the number of comments on the post. Using the post urls, we run a generic news article scraper, used in \cite{horne2018accessing}, to extract article title, body text, and source. We remove any post that has a score of 0 or less, as this means community members have disapproved of the article being posted, which is indicated by ``downvoting" the post to a score of 0.

In addition to this primary data set, for a secondary test of our models over time, we collect and scrape news articles from posts in mainstream and conspiracy during 3 months in 2016 and 3 months in 2015. We choose these two communities for this supplementary test as they both have rich data dating back over 10 years, where as bias1 and bias2 are younger communities.

In total, we analyze over 60K articles across 4 communities. To further illustrate the diversity of these communities, we compute the overlap of news articles posted (Table~\ref{article-overlap}), news sources posted (Table~\ref{source-overlap}), and named entities mentioned in news articles (Table~\ref{entity-overlap}). Specifically, news article overlap is the percent of identical news article posted in a pair of communities, where news source overlap is the percent of articles posted that come from the same source in a pair of communities. Similarly, named entity overlap is the percent of articles that mention the same person, place, or group between a pair of communities. Details about our named entity extraction processes can be found in Section~\ref{feats}. This meta-data shows the largest overlap in news articles is 0.58\% between mainstream and bias2, the largest overlap in sources is 14.1\% between conspiracy and bias2, and the largest overlap in named entities mentioned is 14.0\% between bias1, bias2, and conspiracy. Overall, we see very little similarity between all 4 communities, illustrating a natural separation between the types of news shared in each.

\begin{table} [h]
\begin{center}
\hspace*{-0.0in}\begin{tabular}{|c||c|c|c|c|c|c|c|}
\hline
& \textbf{mainstream} & \textbf{conspiracy} & \textbf{bias1} & \textbf{bias2} \\
\hline
\textbf{mainstream} & \textbf{100\%} & 0.19\% & 0.56\% & 0.58\%\\
\hline
\textbf{conspiracy} & 0.19\% & \textbf{100\%} & 0.53\% & 0.25\% \\
\hline
\textbf{bias1} & 0.56\% & 0.53\% & \textbf{100\%} & 0.37\% \\
\hline
\textbf{bias2} & 0.58\% & 0.25\% & 0.37\% & \textbf{100\%}\\
\hline
\end{tabular}
\caption{Percentage of \texttt{news article} overlap in 2017 data set}\label{article-overlap}
\end{center}
\end{table}

\begin{table} [h]
\begin{center}
\hspace*{-0.0in}\begin{tabular}{|c||c|c|c|c|c|c|c|}
\hline
& \textbf{mainstream} & \textbf{conspiracy} & \textbf{bias1} & \textbf{bias2} \\
\hline
\textbf{mainstream} & \textbf{100\%} & 6.5\% & 10.8\% & 9.6\% \\
\hline
\textbf{conspiracy} & 6.5\% & \textbf{100\%} & 13.6\% & 14.1\% \\
\hline
\textbf{bias1} & 10.8\% & 13.6\% & \textbf{100\%} & 13.7\% \\
\hline
\textbf{bias2} & 9.6\%  & 14.1\%  & 13.7\% & \textbf{100\%}  \\
\hline
\end{tabular}
\caption{Percentage of \texttt{news source} overlap in 2017 data set}\label{source-overlap}
\end{center}
\end{table}

\begin{table}[h]
\begin{center}
\hspace*{-0.0in}\begin{tabular}{|c||c|c|c|c|c|c|c|}
\hline
& \textbf{mainstream} & \textbf{conspiracy} & \textbf{bias1} & \textbf{bias2} \\
\hline
\textbf{mainstream} & \textbf{100\%} & 6.5\% & 11.8\% & 9.6\% \\
\hline
\textbf{conspiracy} & 6.5\% & \textbf{100\%} & 14\% & 14\% \\
\hline
\textbf{bias1} & 11.8\% & 14\% & \textbf{100\%} & 13.6\% \\
\hline
\textbf{bias2} & 9.6\% & 14\% & 13.6\% & \textbf{100\%}  \\
\hline
\end{tabular}
\caption{Percentage of most frequent \texttt{named entity} mentioned overlap in 2017 data set}\label{entity-overlap}
\end{center}
\end{table}

\section{Features}~\label{feats}
In order to capture differences in the news read by each community, we compute 7 groups of features, all of which can be extracted from a news article alone. These feature groups are inspired by the techniques used in~\cite{horne2018accessing}, in addition to, news popularity literature~\cite{bandari2012pulse}. In \cite{horne2018accessing}, a similar set of features is used to predict both the reliability of news and the bias of news with very high accuracy. We expect tight-knit online communities to have preferences that parallel these news article classes to some degree, demonstrating the potential usefulness of these content feature groups. These feature groups are as follows:

\textbf{Style} features are built to capture the overall writing style and structure in a news article. These features include Parts-of-Speech (POS)~\cite{loper2002nltk}, punctuation, use of all capitalized words, use of quotes, use of past, present, or future tense, quantification words~\cite{pennebaker2001linguistic}, and swear words. In total this feature group contains 45 features which are compute on the body text and title text of the news article independently. 

\textbf{Complexity} features capture the complexity of writing in a news article. These features include lexical diversity, reading grade complexity, number of stop words, average word length, and the length of news article. In total this feature group contains 7 features which are compute on the body text and title text of the news article independently. 

\textbf{Bias} features capture how opinionated or one-sided a news story is. These features include bias word count~\cite{ mukherjee2015leveraging, recasens2013linguistic}, number of hedges (i.e. could, maybe, possibly, etc.)~\cite{mukherjee2015leveraging, recasens2013linguistic}, number of factives (a verb, adjective, or noun phrase presupposing the truth of a sentence), number of implicatives (i.e. manage to, failed to, etc.), certainty/tentativeness~\cite{pennebaker2001linguistic}, and subjectivity~\cite{horne2018sampling}. In total this feature group contains 11 features which are compute on the body text and title text of the news article independently.

\textbf{Named Entity} features capture who and what is being talked about in a news article. Specifically, we extract the most frequently mentioned named entity from each news article and encode it into a unique number. Examples of named entities include: Steven Hawking (person), Middle East (place), ISIS (group), and Illuminati (group). Named entity extraction is done using Python NLTK~\cite{loper2002nltk}.

\textbf{Sentiment} features capture the emotion and affect in a news article. These features include positive emotion words~\cite{hutto2014vader, pennebaker2001linguistic}, negative emotion words, neutral emotion words, words that indicate anger, words that indicate assent~\cite{pennebaker2001linguistic}, and the strength of those words~\cite{recasens2013linguistic}. In total this feature group contains 16 features which are compute on the body text and title text of the news article independently.

\textbf{Entity Slant} is a combination of our sentiment feature group and entity feature group. This is a simple method of capturing the affect towards a named entity in a news article. While this feature could be much more granular, such as being computed per sentence rather than per article, it should capture the relative slant towards or against the most frequently mentioned entity. In total this feature group contains 17 features.

\textbf{Source} simply captures what news sources a community prefers to read. To do this, we build a source encoding dictionary that assigns a unique number to each source. If two communities read from mutually exclusive sets of sources, this feature will completely separate the communities. 

\section{Machine Learning Models}
Using these feature groups, we implement several well-known machine learning algorithms to test which features are best for prediction. Specifically, we use linear-kernel Support Vector Machines (SVM) and Random Forest (RF) classifiers. Each algorithm's hyper-parameters are tuned using 10-fold cross validation. Further, each model is trained using balanced class weights. The classes are natural imbalanced due to varying posting behavior in each community. While this imbalance is not very extreme, it is best practice to train with balanced classes. This balance could also be achieved using minor-class oversampling or SMOTE balancing, which is a synthetic data point technique. These techniques are typically used when the imbalance in classes is extreme. All algorithms are implemented using Python Sci-kit Learn~\cite{pedregosa2011scikit}.

\begin{table*}
\begin{center}
\hspace*{-0.0in}\begin{tabular}{|cccc|}
\hline
& \textbf{conspiracy} & \textbf{bias1} & \textbf{bias2}  \\
\textbf{mainstream} & \raisebox{-.5\height}{\includegraphics[width=5cm]{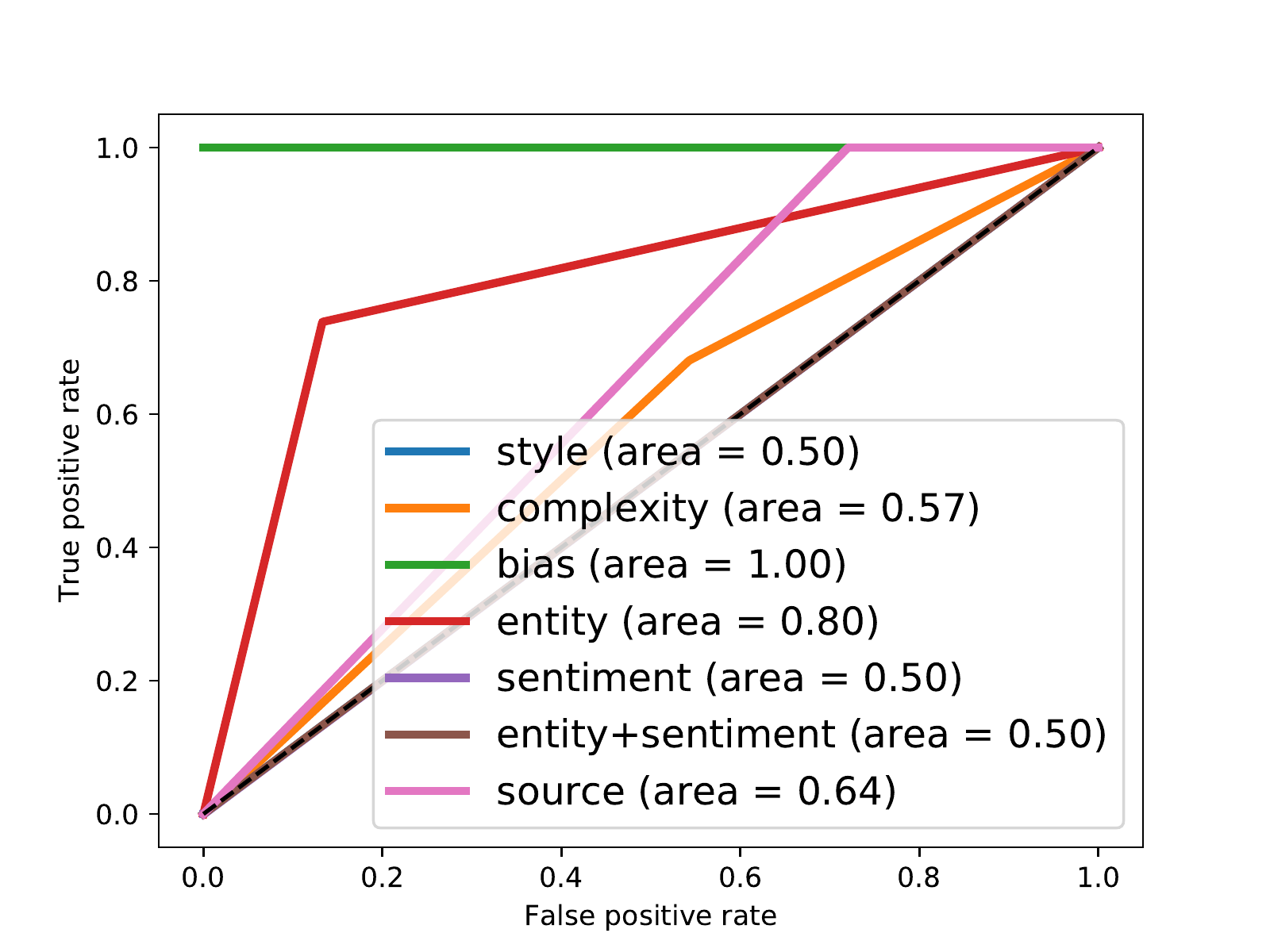}} & \raisebox{-.5\height}{\includegraphics[width=5cm]{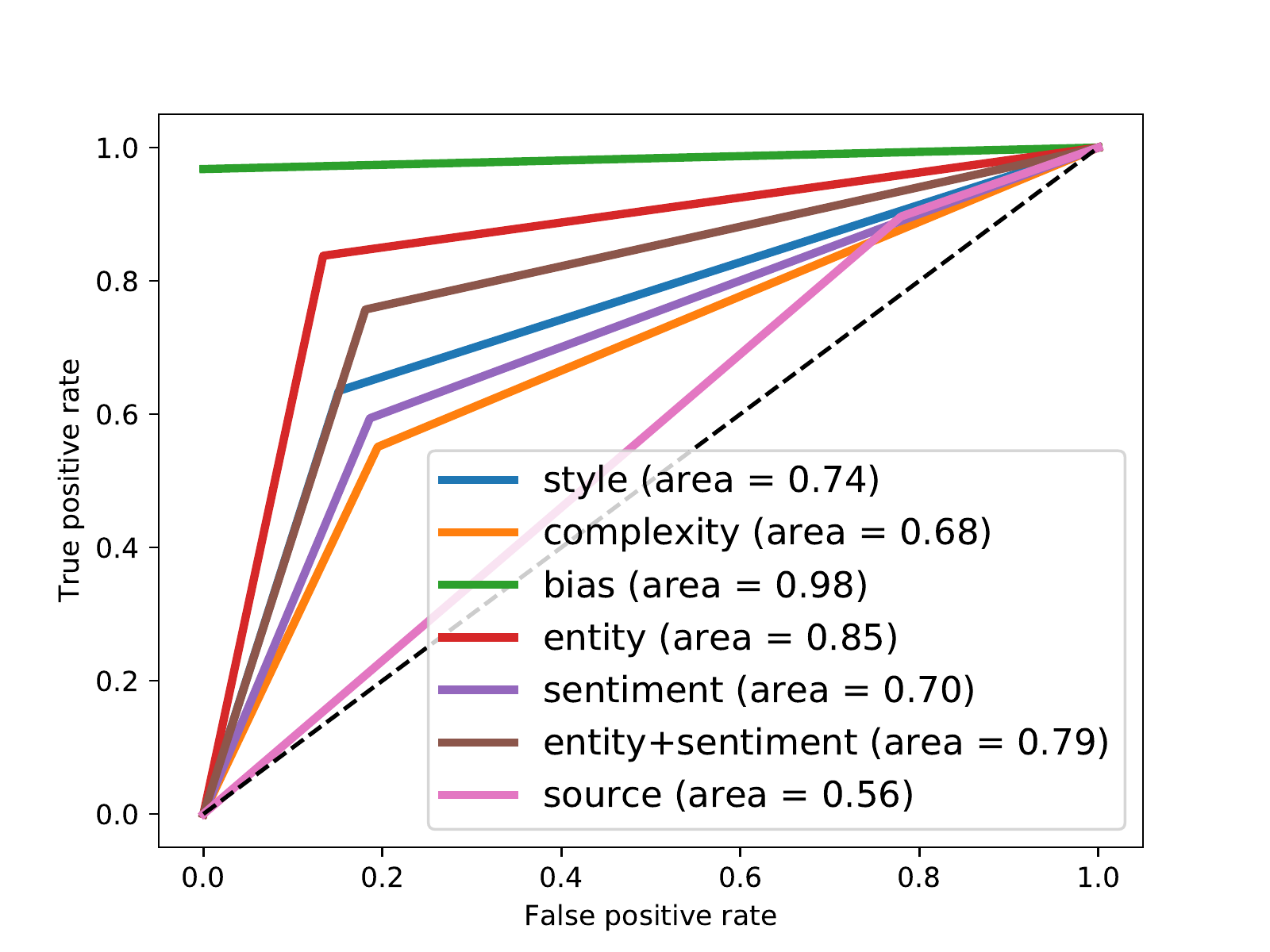}} &  \raisebox{-.5\height}{\includegraphics[width=5cm]{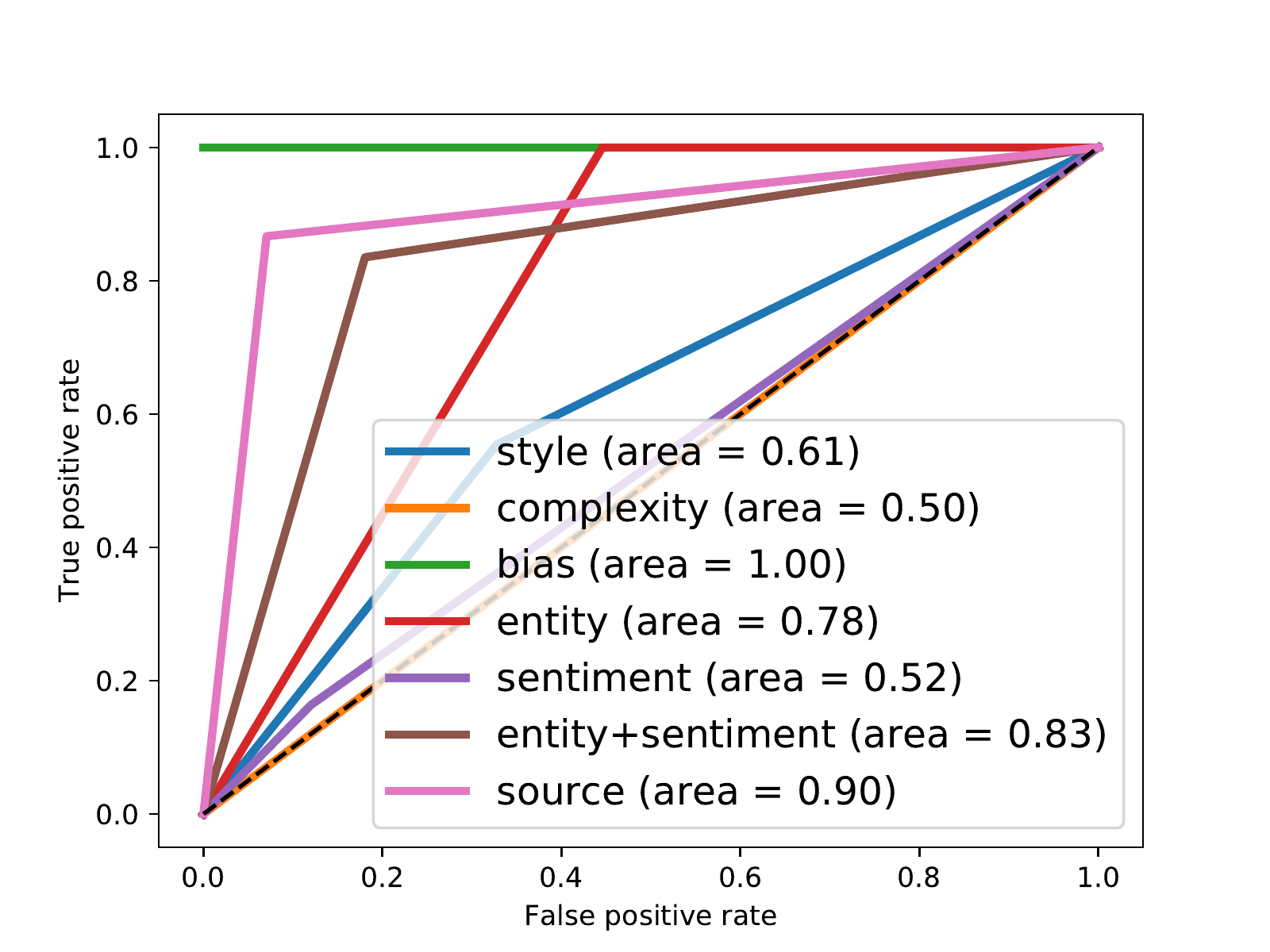}}\\

\textbf{conspiracy}  & \raisebox{-.5\height}{\includegraphics[width=5cm]{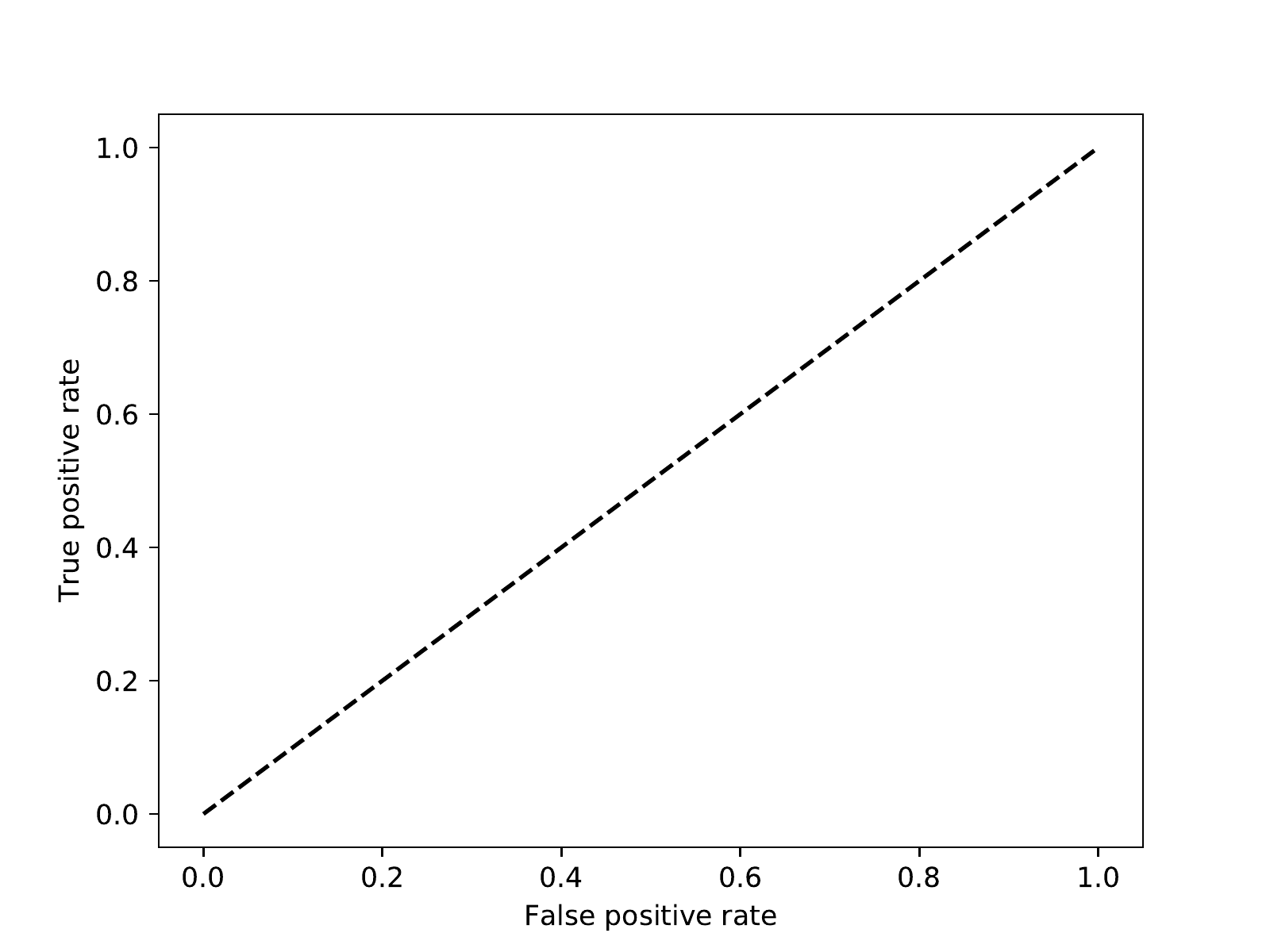}} &  \raisebox{-.5\height}{\includegraphics[width=5cm]{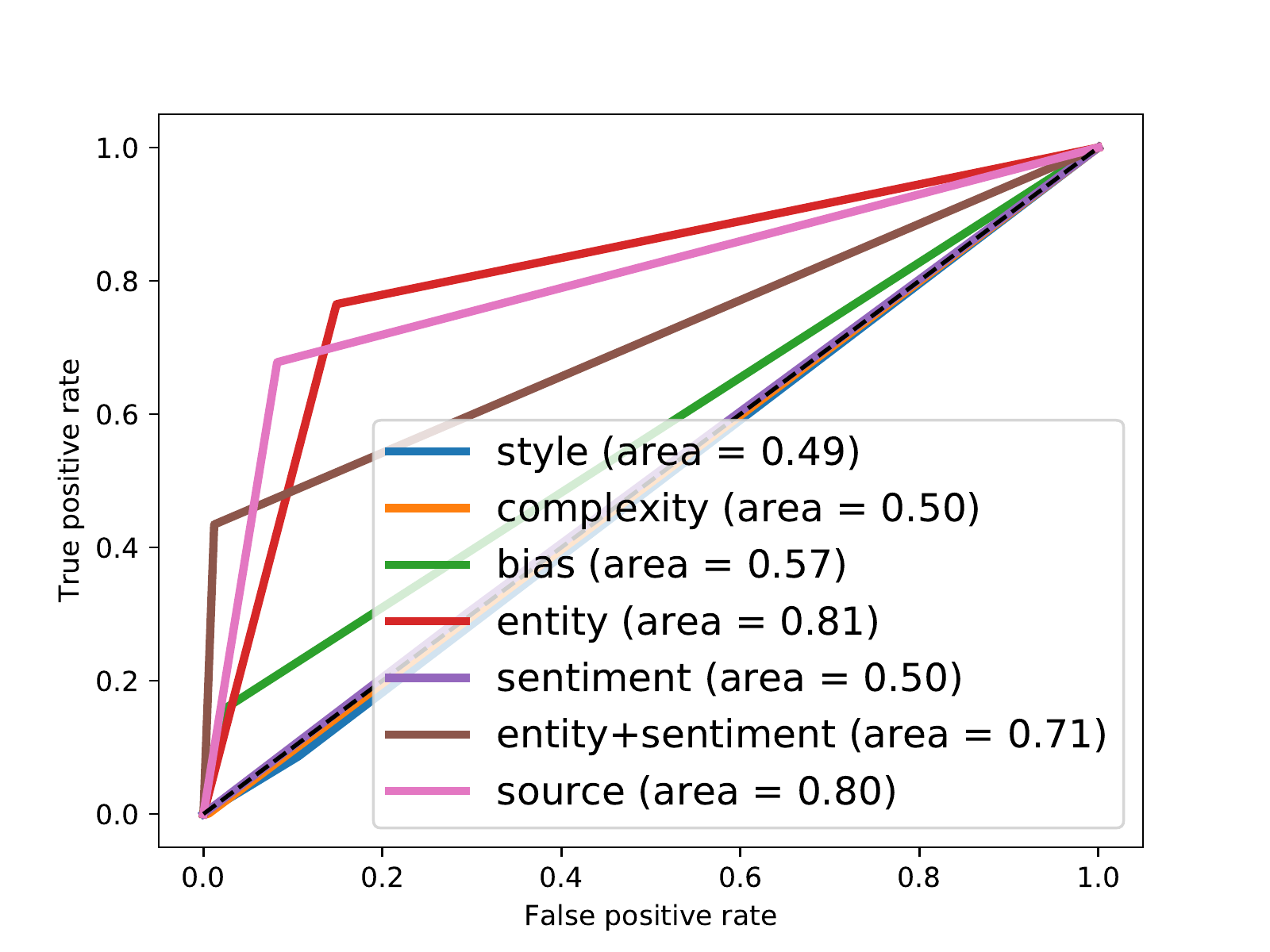}}& \raisebox{-.5\height}{\includegraphics[width=5cm]{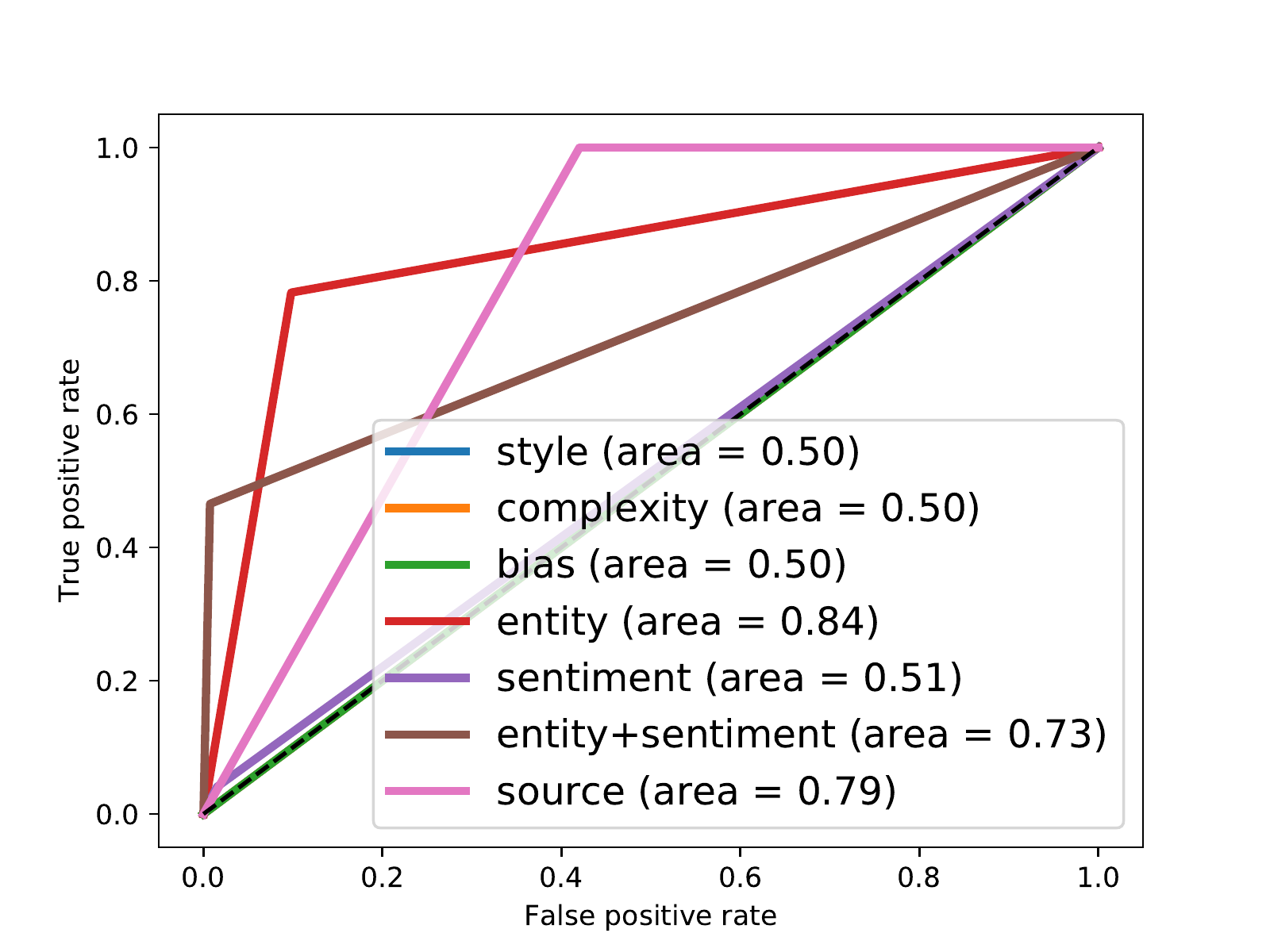}} \\

\textbf{bias1}  &  & \raisebox{-.5\height}{\includegraphics[width=5cm]{Figures/ROCAUC_Blank.pdf}} & \raisebox{-.5\height}{\includegraphics[width=5cm]{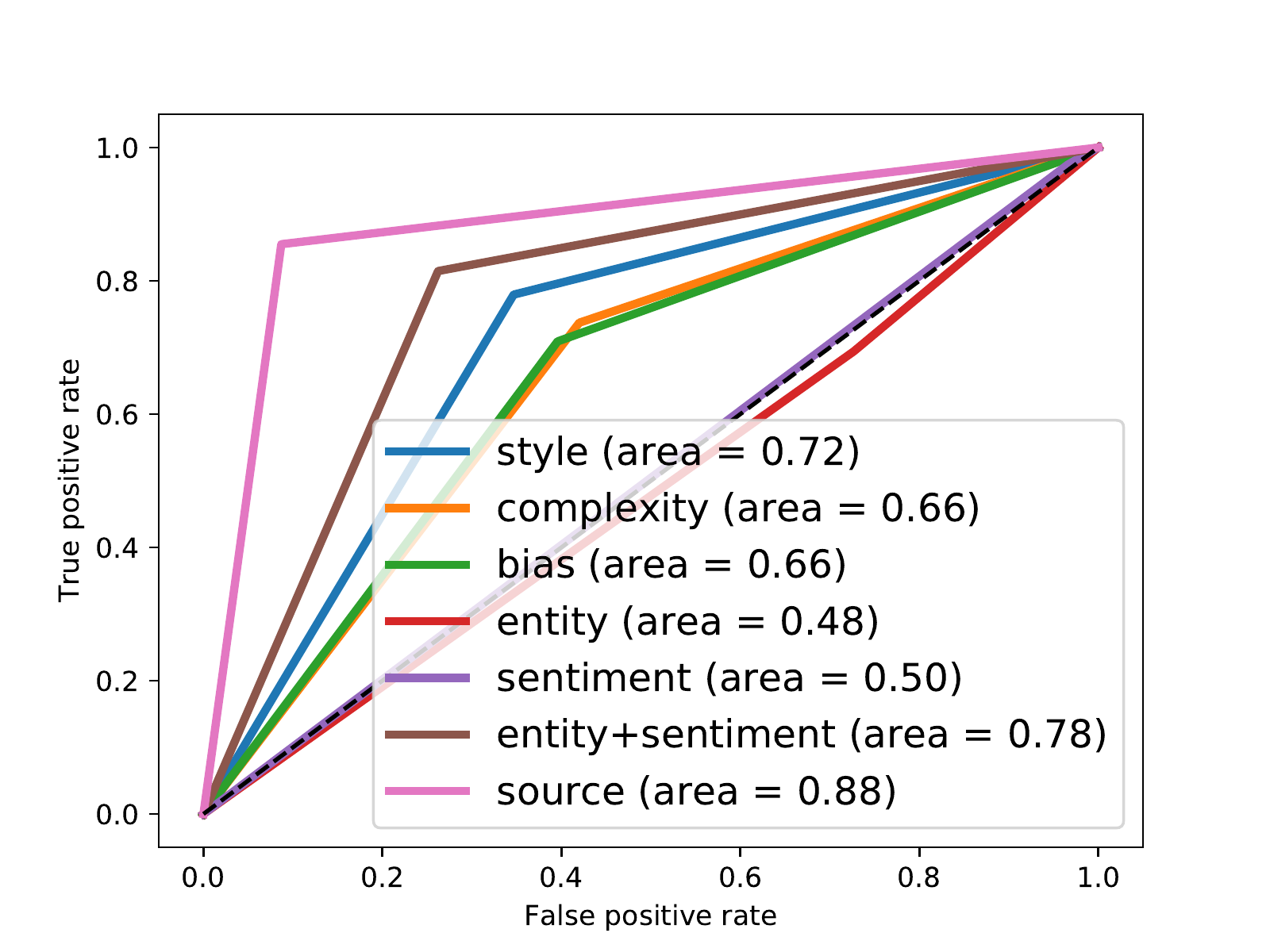}} \\
\hline
\end{tabular}
\caption{Confusion Matrix of ROC curves for each Random Forest model. Each cell in the table contains the ROC curves (and AUC) of each feature group for classifying news articles into the respective communities. For example, row 1, column 1 contains the ROC graph for classifying news articles between mainstream and conspiracy. To save space, we only show the top-half of the matrix, as it is symmetrically on the diagonal.}\label{results}
\end{center}
\end{table*}

 \begin{table*}[ht]
\centering
\hspace*{-0.4in}\begin{tabular}{cccc}
 \small{(a) Train/Test 2017} & \hspace*{-0.25in}\small{(b) Train/Test 2016} & \hspace*{-0.25in}\small{(c) Train/Test 2015} & \hspace*{-0.25in}\small{(d) Test Over Time}\\
\includegraphics[width=5.2cm]{Figures/ROCAUC_news-conspiracy-allarticles.pdf}&
\hspace*{-0.25in}\includegraphics[width=5.2cm]{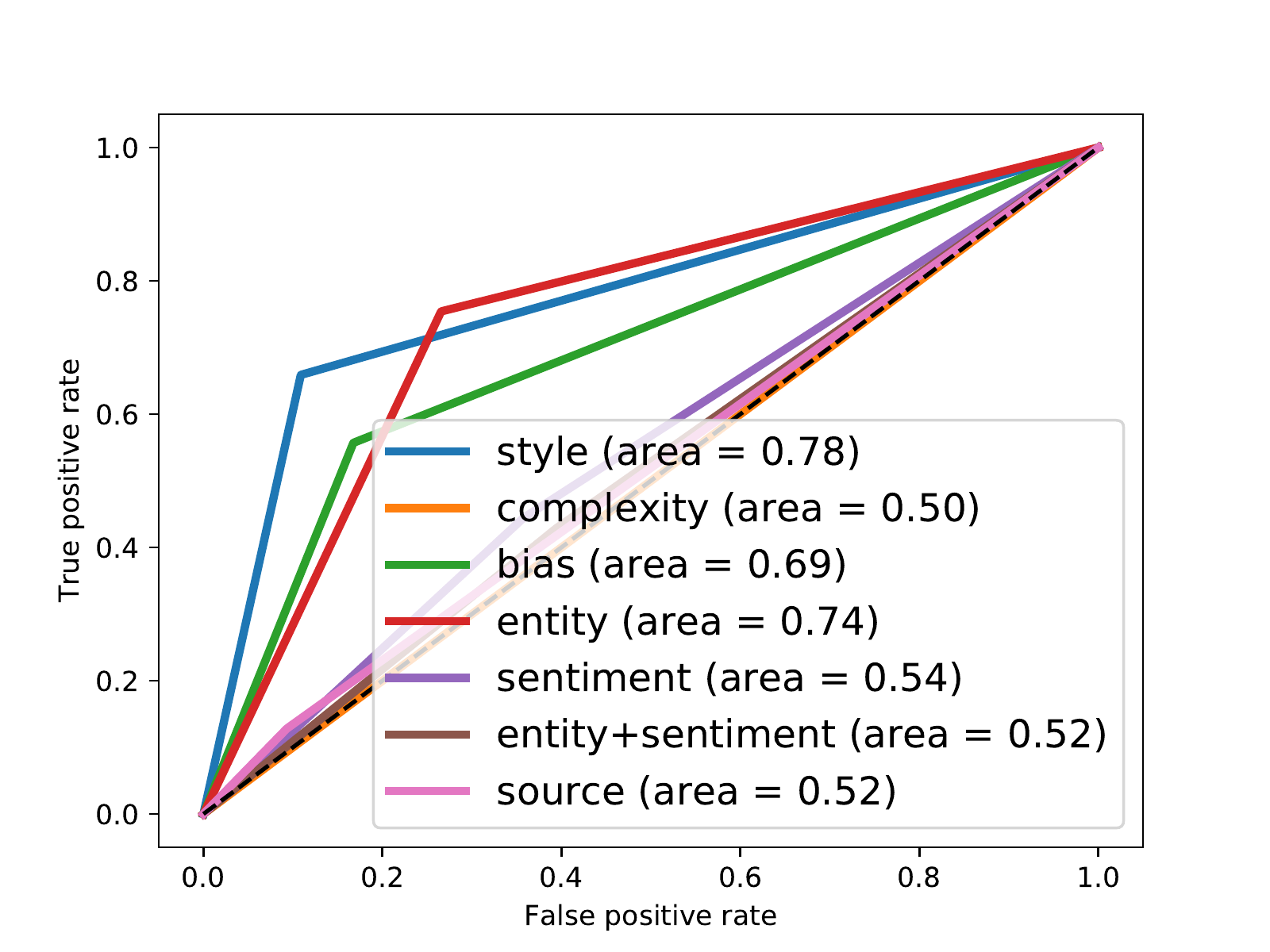}&
\hspace*{-0.25in}\includegraphics[width=5.2cm]{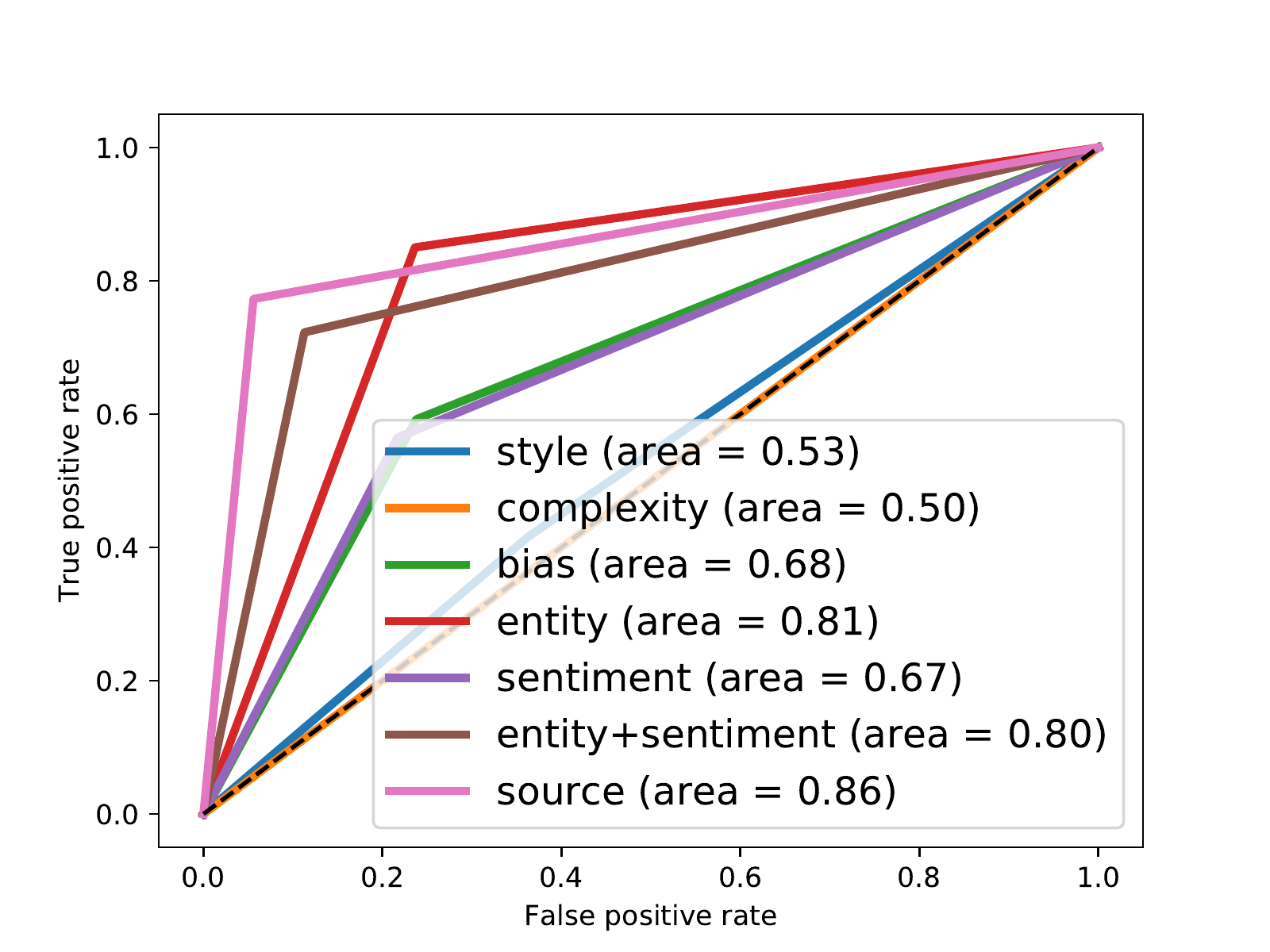}&
\hspace*{-0.25in}\includegraphics[width=5.2cm]{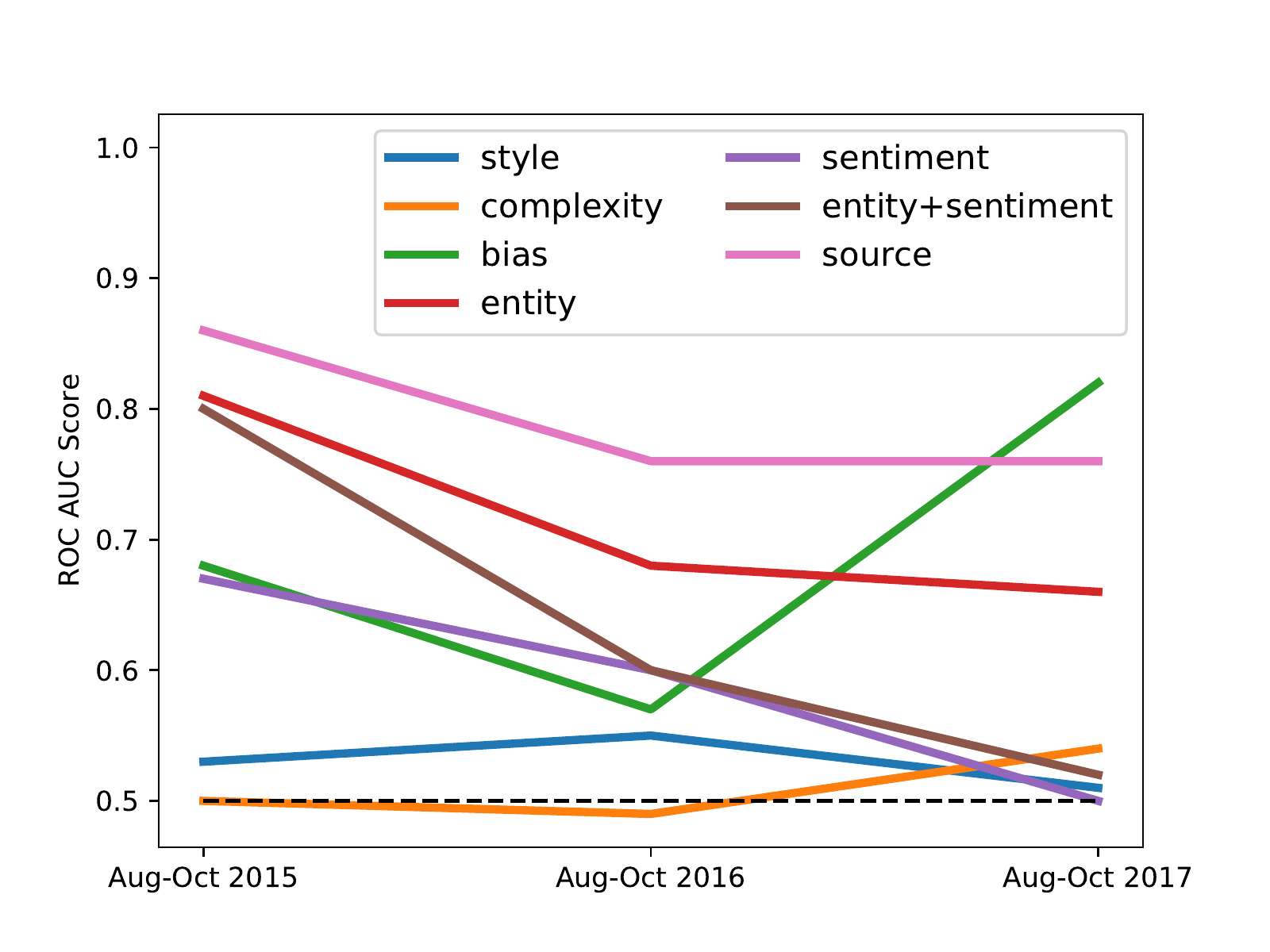}\\
\end{tabular}
\caption{ROC AUC scores for each feature group over time when classifying \textbf{mainstream} vs \textbf{conspiracy}. Specifically: \textbf{(a)} Training and testing each model on 2017 data, \textbf{(b)} Training and testing each model on 2016 data, \textbf{(c)} Training and testing each model on 2015 data, and \textbf{(d)} Training models with 2015 data, testing on 2016 and 2017 data.}
  \label{time}
\end{table*} 

\section{Results}
First, we examine how well each of our models perform on the 2017 data set. Due to limited space, we only show results for our Random Forest models. However, we found very similar results using linear-kernel SVMs. In addition, we try various thresholds of popularity using both the voting score and the number of comments (for example, training on the top 20\% of articles by score, top 30\% by score, etc.), but find little difference in performance. In fact, we find that when we use all posts, with the exception of posts with a score of 0, our performance is slightly better. Therefore, the results we show are using all posts with scores above 0 for training and testing.

A confusion matrix of Receiver Operating Characteristic (ROC) curves for each community pair and feature group can be found in Table~\ref{results}. Specifically, we train and test classification between each binary pair of communities for each feature group. This allows us to understand exactly what features provide clear signal and how those signals differ between communities. In each graph's legend, the ROC Area Under The Curve (AUC) values can be found. As a rule of thumb, ROC AUC values can be interpreted as such:  0.90 to 1.0 is excellent, 0.80 to 0.90 is good, 0.70 to 0.80 is fair, 0.60 to 0.70 is poor, and 0.50 to 0.60 is fail. In each graph, we plot a black dotted line to indicate a ROC AUC of 0.5, which is random chance.

\subsection{Models for Prediction}
\textbf{We can predict which community is interested in a news article.} In Table~\ref{results}, we see each community pair can be classified with reasonable accuracy. The community mainstream can be separated from each other community the best, achieving near 1.0 AUC. While separating conspiracy from bias1 achieves 0.81 AUC at its best and conspiracy from bias2 achieves 0.84 AUC at its best.

\textbf{These prediction models are community-pair dependent.} While each community pair can be separated, very different sets of features are used to classify. Features that best differentiate mainstream articles from conspiracy articles are bias, entity, and source. Features that differentiate mainstream articles from bias1 articles are bias, entity, sentiment, entity slant, and style. Similarly, features that differentiate mainstream articles from bias2 articles are bias, entity, entity slant, and source. Bias and entity based features clearly separate mainstream news communities from alternative news communities. On the other hand, features that separate bias1 from bias2 are source and entity slant. Interestingly, we see that entity features on their own do very poorly (0.48 AUC), but entity slant does well (0.78 AUC). This shows that the hyper-partisan communities are talking about the same people, places, and things, but with a different affect towards them, as we naturally expect. Lastly, we see that conspiracy articles only are separated from bias1 and bias2 articles with source and entity features.  

\subsection{Generalizing Models Over Time}
An important metric of performance for machine learning models is how well they work over long periods of time. This notion is often called ``concept drift," which refers to unforeseen changes in a target variable over time~\cite{vzliobaite2010learning}. Concept drift becomes particularly important when prediction models are applied to quickly evolving situations, such as predicting social concepts, the news cycle, or fraud detection~\cite{dal2015credit}. While a model can perform very well in a small time frame, its performance may degrade over time. To test this, we train classifiers using the 2015 data to predict news article interest in 2016 data and 2017 data. We only run this test for mainstream and conspiracy, as they have been very active communities for a long period of time, allowing us to maintain a large and rich data set in all 3 years. Further, we train and test new models on the 2016 data and 2015 data to show performance within those time-frames. These results can be found in Table~\ref{time}.

\textbf{Some feature groups can generalize over time, others cannot.} In Table~\ref{time}(d), we show the performance of each feature group when the model is trained on 2015 data and tested on test sets from all three time frames (2015, 2016, 2017). This test is meant to emulate the performance of a 2015 trained model that is not retrained for 3 years. In general, we see small decreases in performance for each feature group as time moves further away from the trained model. Overall, due to the ever-changing news cycle, this finding is expected. However, these decreases in performance vary for each feature group. The largest decreases in performance come from entity slant and sentiment features, where as the source and entity features have a less significant drop. Considering the large amount of time between each data set, the drop in performance for source features is not very significant, dropping from 0.86 AUC to 0.76 AUC. Interestingly, while the performance for our bias features decreases in the first year, it significantly increases in the second year. This drastic change in performance may show a shift in news producer or news community behavior between 2016 and 2017. Specifically, the model's knowledge of bias behavior remains in the same direction (conspiracy interested articles being more bias than mainstream interested articles), but the separation between the classes increases, allowing the model to make less mistakes using the 2015 decision boundary. This shift could be consider a phase transition, but more research is certainly required to understand exactly why that transition occurs and what effects this transition may have.  

\textbf{Retraining the models over time can prevent the degrade in performance.} Despite certain feature models breakdown in performance, we do see clear signal in the models when they are trained on data from the same time frame. Table~\ref{time}(a), \ref{time}(b), and \ref{time}(c) show the ROC curves for each feature group in each time frame. We see that while performance and feature importance differ across the time frames, each one shows fair to excellent performance, illustrating models can be retrained to maintain performance over time. How often a model should be retrained will depend on the accuracy tolerance for a particular application, and could be determined with more granular data.

\section{Conclusions and Discussion}
In this paper, we examine community-specific interest in news articles. We construct supervised machine learning models to predict which community will be most interested in a news article using distinct communities on Reddit. Additionally, we assess the concept drift effects on each feature group over a 3 year time frame. Our results show that we can predict community interest with high accuracy, but these models are community dependent, feature group dependent, and can degrade over-time.

These results reveal several strategies to better tackle community-specific interest predictions. First, the strongly community dependent nature of the feature group importance suggests that \textit{hierarchical-binary} models should be used over standard \textit{multi-class} models. For example, looking at our 2017 results, we see mainstream articles can be separated from all other communities using bias features and entity features, but bias1, bias2, and conspiracy can only be separated by style, source, and entity-slant features. Hence, a prediction model can be built to first classify a news article as \textit{mainstream} or \textit{not} using bias and entity features. If that article is classified as \textit{not mainstream}, the article can be passed through several binary classifiers which use different feature groups to classify the article into bias1, bias2, or conspiracy. On the other hand, if we used a multi-class model, we would have to include all feature groups and tune the model to all communities, which would likely underfit the data, leading to very poor performance.

Second, models should be critically analyzed for performance loss over time. The change in performance we found in this study illustrates the inherent complexity of the problem. With more granular data, we can learn how often models need to be retrained in order to maintain a certain level of performance. For example, if we test our models on week-long intervals of data, we can determine when the performance loss has degraded too much according to a tolerance threshold. At this threshold point, we can automatically retrain the models. This pattern of performance loss is likely very dependent on the communities themselves, as some may evolve slower than others. Further, performance loss may not be a steady decline, as shown by our bias feature model, which actually has improved performance over the 3 year time frame, suggesting concept drift in these models can be very complex. From a more general point of view, these results suggest that all machine learning models should be put through a time test to ensure desired performance levels are met.

\section{Acknowledgment}
\small{This work was partially supported by the U.S. Army Research Laboratory under Cooperative Agreement No. W911NF-09-2-0053; The views and conclusions contained in this document are those of the authors and should not be interpreted as representing the official policies, either expressed or implied, of ARL, NSF, or the U.S. Government. The U.S. Government is authorized to reproduce and distribute reprints for Government purposes notwithstanding any copyright notation here on. This document does not contain technology or technical data controlled under either the U.S. International Traffic in Arms Regulations or the U.S. Export Administration Regulations.}

\bibliographystyle{IEEETranSN}
{\small
\bibliography{references}
}
\end{document}